\documentclass[aps,twocolumn]{revtex4}%
\usepackage{amsfonts}
\usepackage{amsmath}
\usepackage{amssymb}
\usepackage{graphicx}%
\begin{document}
\title{Measurement of the inner horizon in the analog of rotating BTZ black holes by
an improved photon fluid }
\author{Siyao Wu}
\affiliation{School of Mathematics and Physics, China University of Geosciences, Wuhan
430074, China}
\author{Ling Chen}
\email{lingchen@cug.edu.cn}
\affiliation{School of Mathematics and Physics, China University of Geosciences, Wuhan
430074, China}
\author{Bolong Yi}
\affiliation{School of Mathematics and Physics, China University of Geosciences, Wuhan
430074, China}
\author{Lei Li}
\affiliation{School of Mathematics and Physics, China University of Geosciences, Wuhan
430074, China}
\author{Baocheng Zhang}
\email{zhangbaocheng@cug.edu.cn}
\affiliation{School of Mathematics and Physics, China University of Geosciences, Wuhan
430074, China}
\pacs{PACS number}

\begin{abstract}
We study how to include the inner horizon in the analog of rotating black
holes using photon fluids. We find that a vortex beam carrying an improved
phase can simulate the rotating BTZ black holes experimentally. In the
experiment, we develop a new photon fluid model in a graphene/methanol thermal
optical solution, and measure the variation of photon fluid velocity with the
radial position using a Fourier plane light spot localization method, while
also determining the variation of phonon velocity with the same radial
position from the optical vortex intensity distribution. The result provides
an extension for the application of optical vortex and a potential possibility
for the future experimental exploration about the properties of BTZ black
holes and even the anti-de Sitter space.

\end{abstract}
\maketitle

\section{Introduction}

Photon fluids \cite{cc2013}, capable of carrying angular momentum, are ideal
candidates for analogue gravity and have been used to simulate the rotating
black holes \cite{vmp2018}. This simulation is achieved at room temperature
through the interaction of vortex light with a self-defocusing medium. The
self-defocusing medium \cite{fm2008} has a nonlinear refractive index directly
related to light intensity, which causes the refractive index to exhibit local
negative curvature as light propagates through it, thereby influencing the
beam itself. From a microscopic perspective, this effect can be understood as
a repulsive interaction between photons mediated by atoms, leading to the
formation of a \textquotedblleft photon fluid\textquotedblright\ state
\cite{fpr1992,cb1999,vwf2016}. Moreover, the propagation of linear excitations
(sound waves) in this fluid occurs as if in an effectively curved spacetime
determined by the physical properties of the photon fluid. This analogous
curved spacetime can be stationary (e.g., a rotating black hole) due to the
stability of the vortex with radial flow and the simultaneous existence of the
radial and angular flow, whose stability can be ensured by the nonlocally
thermo-optic nonlinearity \cite{gct2007,vrf2015}.

The simulation of a curved spacetime can be traced back to the thought of the
analogue gravity put forward in the year of 1981 \cite{wgu1981,blv2011}. Up to
now, it has been realized experimentally in many different physical systems,
such as light in a nonlinear medium \cite{pkl2008,drl2019}, Bose--Einstein
condensates (BEC) \cite{lis2010,ngs2019}, moving water \cite{wtl2011},
superconducting quantum system \cite{syf2023}, and others. It is worth noting
that these analogue systems have typically been limited to (1+1)-dimensional
spacetime, often modeling simple structures like Schwarzschild black holes.
Along the earlier studies \cite{ybz1972,tpw2017} in which the physical
phenomena occurring in the rotating spacetime were simulated, Vocke et al.
\cite{vmp2018} presented experimentally a (2+1)-dimensional rotating spacetime
with the definite identification of the ergosphere besides the horizon. In
their experiment, the rotating analogue black hole was realized by an optical
vortex propagating through a self-defocusing medium at room temperature
\cite{fm2008,vmp2018}, where an ergosurface formed, enclosing the internal
region from which rotational energy could be extracted \cite{bfw2020}.

Since earlier models \cite{vmp2018,fm2008} could not include the inner horizon
(it is noted that the inner horizon was mentioned in the earlier experiment
using BEC to simulate the black hole \cite{js2014}, but it is essentially
different from ours since there is no rotation for the earlier model) and only
represented the equatorial slice of the Kerr geometry, it was proposed
theoretically \cite{czz2023} that a new type of optical vortex should be used
to simulate the Ba\~{n}ado-Teitelboim-Zanelli (BTZ) black hole spacetime
\cite{btz1992}. The BTZ black hole is a solution to Einstein's field equations
in (2+1)-dimensional gravity with a negative cosmological constant. It has
some key differences from Schwarzschild or Kerr black holes \cite{bhtz1993};
for instance, it is asymptotically anti-de Sitter rather than asymptotically
flat and lacks a curvature singularity at the origin. This makes it a physical
platform for exploring the properties of AdS black holes \cite{sc1995},
including the thermodynamics, phase transition, etc, when the radiations can
be observed in the future. Further, the AdS/CFT correspondence
\cite{agmo2000,ae2015} had been investigated theoretically in the context of
analog gravity \cite{sh2015,bdt2015,gsz2015,dlt2016,gkg2018}. In particular,
different types of AdS black holes could correspond to different gravitational
theories or CFTs, such as the exotic BTZ black holes \cite{tz2013,zbc2013} and
planar black holes \cite{sh2016}. Therefore, experimentally realizing these
black holes is crucial for understanding the theory itself and offers further
possibilities for exploring them. In this paper, we experimentally realize a
class of AdS black holes, namely the BTZ black holes, using vortex beams,
which is highly significant not only for the extensive application of the
optical vortex, but also for further exploring the AdS/CFT theory in the laboratory.

The structure of the paper is organized as follows. In Section 2, we describe
the analogue model of the BTZ black hole and the necessary optical vortex. In
Section 3, we present the experimental results and related details for the
analogue BTZ black hole. Finally, in Section 4, we give the conclusion.

\section{The model}

In order to simulate a rotating acoustic BTZ black hole, there should be an
effective method to form the inner horizon beside that realized in the earlier
simulation for the rotating black holes where the event horizon and the
ergosphere had been formed. We consider a monochromatic vortex beam with the
topological charge $m$ propagating through a nonlinear medium such as a dilute
methanol/graphene solution. The evolution of the photon fluid, when
interacting with the medium, is governed by the nonlinear Schr\"{o}dinger
equation (NLSE), in which the thermo-optic nonlinearity is represented by the
nonlinear change in refractive index, as expressed in \cite{vmp2018,vrf2015}%
\begin{equation}
\Delta n=\gamma\int R\left(  r-r^{\prime},z-z^{\prime}\right)  I(r^{\prime
},z^{\prime})dr^{\prime}dz^{\prime}, \label{index}%
\end{equation}
where $\gamma$ is the nonlinearity coefficient, $z$ is the propagation
direction, $I$ is the light intensity, and $R(r,z)$ is the response function,
which depends on the nonlocal processes within the medium. When heat diffusion
is limited, the nonlocal response function can be approximated as a delta
function, a valid assumption in our study, as discussed in Ref. \cite{vmp2018}%
. In partiuclar, the properties of photon fluid formed in the interaction
between the light and the medium in the solution is also controlled by the
phase and intensity of the incident optical field. The superfluid behavior of
photon fluid is crucial for our experiment, which requires that the wavelength
of sound modes simulated in the nonlinear medium is much larger than the
healing length. Thus, a linear dispersion for these sound modes is guaranteed.
At the same time, the healing length cannot too small to ensure the stability
of the fluid. So the light has to propagate in the nonlinear medium for at
least one oscillation period to guarantee all those requirements mentioned
above, which is realized in our experiment. The detailed discussion see Ref.
\cite{vmp2018,vrf2015}.

As in the earlier analyses \cite{vmp2018,fm2008,czz2023}, the photon fluid has
a lateral speed gradient distribution which leads to an analogue spacetime
structure with the metric as%
\begin{align}
\mathrm{d}s^{2}  &  =(\frac{\rho_{0}}{c_{s}})^{2}[-(c_{s}^{2}-v_{t}%
^{2})\mathrm{d}t^{2}-2v_{r}\mathrm{d}r\mathrm{d}t\nonumber\\
&  -2v_{\theta}r\mathrm{d}\theta\mathrm{d}t+dr^{2}+(r\mathrm{d}\theta)^{2}],
\label{obhm}%
\end{align}
where $c_{s}$ is the local speed of sound and is related to the bulk pressure
which is derived from the nonlinear interaction. $c_{s}=\frac{c}{n_{0}}%
\sqrt{\frac{\left\vert \gamma\right\vert I}{n_{0}}}$ where $c$ is the light
velocity in the vacuum, $n_{0}$ is the linear refractive index of the sample
solution, $\gamma$ is the nonlinearity coefficient, and $I$ is the light
intensity within the region selected by the pinhole. The terms $v_{r}$ and
$v_{\theta}$ represent the radial and tangential velocity components, from
which the total velocity is $v_{t}^{2}=v_{r}^{2}+v_{\theta}^{2}$. For the
metric (\ref{obhm}), it can model a spacetime about a black hole if the region
$v_{r}>c_{s}$ exists where anything cannot escape, and it can model a
spacetime about the ergosurface of a rotating black hole if the region
$v_{t}>c_{s}>v_{r}$ exists where it is impossible for an observer to remain
stationary relative to a distant observer. So, it is appropriate to simulate
the properties of a rotating black hole, as made in Ref. \cite{vmp2018}, but
the inner horizon can not be found in their work. Here, we firstly see if the
metric (\ref{obhm}) can match the BTZ black hole metric in form.

It is not hard to get that the BTZ black hole metric can be simulated by
letting \cite{czz2023}%

\begin{equation}
v_{r}=c_{s}\sqrt{1-(-M+\frac{r^{2}}{l^{2}}+\frac{J^{2}}{4r^{2}})}, \label{frv}%
\end{equation}

\begin{equation}
v_{\theta}=c_{s}\frac{J}{2r}, \label{fav}%
\end{equation}
up to the overall factor $(\frac{\rho_{0}}{c_{s}})^{2}$. Here the constants
$G=c=1$ are taken. For BTZ black holes, $\Lambda=-\frac{1}{l^{2}}$ is the
negative cosmological constant and $l$ is the radius of the anti-de Sitter
space. $M$ and $J$ are the mass and angular momentum of the black hole. The
positions of the outer ($r_{+}$) and inner ($r_{-}$) horizon of the black hole
are determined by $-M+\frac{r^{2}}{l^{2}}+\frac{J^{2}}{4r^{2}}=0$, which leads
to the results, $r_{\pm}^{2}=\frac{Ml^{2}}{2}\left(  1\pm\sqrt{1-(\frac{J}%
{Ml})^{2}}\right)  $. We may assume without loss of generality that $J\geq0$
and assume that $Ml\geq J$ to ensure the existence of an event horizon at
$r=r_{+}$. The ergosurface is specified at the position $r_{E}=\sqrt{M}%
l=\sqrt{r_{+}^{2}+r_{-}^{2}}$. In particular, the metric (\ref{obhm}) is valid
only in the range where the radial velocity $v_{r}=c_{s}\sqrt{1-(-M+\frac
{r^{2}}{l^{2}}+\frac{J^{2}}{4r^{2}})}$ is well defined, and this holds for
$R_{-}<r_{-}<r_{+}<R_{+}$ with%

\begin{equation}
R_{\pm}^{2}=\frac{(1+M)l^{2}}{2}\left[  1\pm\sqrt{1-(\frac{J}{(1+M)l})^{2}%
}\right]  , \label{fbhb}%
\end{equation}
where $R_{\pm}$ represent the boundary that the analogue can be made.
Therefore, the acoustic metric realized by the optical vortex can be regarded
as the analogue of the BTZ black hole with the nearly perfect match in form up
to an overall factor.

\begin{figure}[ptb]
\centering
\includegraphics[width=0.9\columnwidth]{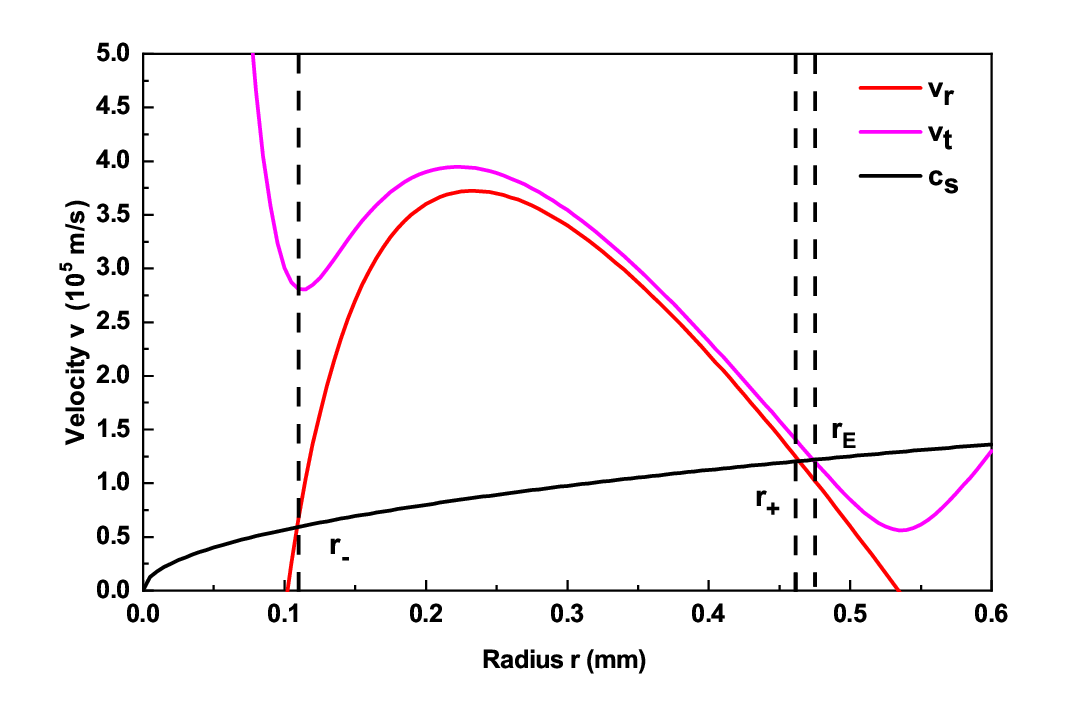} \caption{(Color online) The
velocities of the fluid as a function of the radius. The parameters are taken
according to the actual experimental requirements, i.e. the linear refractive
index $n_{0}=1.33$, the nonlinear coefficient $|\gamma|=4.4\times10^{-11}
m^{2}/W$, the power of the laser $p=140$ mW, the beam waist of the laser
$\sigma=5$ mm, and the topological charge integer $m=2$.}%
\label{Fig1}%
\end{figure}

Then, it is crucial that whether there exists a proper optical vortex which
can realize the BTZ spacetime structure. This has been given in Ref.
\cite{czz2023} by a special Gauss optical vortex with the electric field
\begin{equation}
E_{0}=\sqrt{\rho_{0}(r)}\mathrm{exp}\left(  \mathrm{i}m\theta-2\mathrm{i}%
\pi\left(  Ar-B\mathrm{ln}r-Cr^{2}\right)  \right)  , \label{novf}%
\end{equation}
where $\rho_{0}(r)=\rho_{0}\mathrm{\exp}\left(  -\frac{r^{2}}{\sigma^{2}%
}\right)  \tanh^{m/4}(\frac{r}{\sigma})$ takes the super-Gaussian vortex form
and $\rho_{0}$, $A$, $B$, $C$ are the constants, and the phase
\begin{equation}
\varphi\left(  r\right)  =m\theta-2\pi(Ar-B\mathrm{ln}r-Cr^{2}),
\end{equation}
which gives the radial and tangential velocities as $v_{r}=\frac{-2\pi
c}{kn_{0}}\left(  A-\frac{B}{r}-2Cr\right)  $ and $v_{\theta}=\frac{cm}%
{kn_{0}r}$. In the model analogous to the BTZ black hole, the parameters $A$,
$B$, and $C$ are subject to specific constraints and they cannot be
arbitrarily assigned values to resemble the BTZ black hole. All three
parameters must be positive. To ensure the accuracy of the model, it is
necessary to precisely adjust the relevant parameters based on the actual
structure of the BTZ black hole (as shown in Fig. 1). Specifically, it is
essential to ensure that the curves of radial flow velocity and local sound
speed have two distinct intersection points, which correspond to the inner and
outer horizons of the black hole, respectively. Additionally, the curve of
total flow velocity and local sound speed should also have an intersection
point, which represents the static boundary. Furthermore, the specific
locations of these three intersection points must satisfy certain conditions,
i.e. $R_{-}<r_{-}<$ $r_{+}<$ $r_{E}<R_{+}$. Here we take $A=14/2\pi$,
$B=1.2/2\pi$, $C=22/2\pi$, and $m=2$ for the topological charge of the optical
vortex. The theoretical result for the analogue BTZ black hole is presented in
Fig. 1 where the inner and outer horizon and the ergosurface is seen clearly.

\section{Experimental results}

The schematic diagram of the experimental setup is shown in Fig. 2. In the
experiment, a continuous wave laser source with a super-Gaussian distribution
and a vacuum wavelength of $\lambda=532$ nm is used to generate vortex light.
The laser has a maximum power of up to $200$ mW. Initially, the beam is
adjusted by passing through a beam expander to increase its diameter. The
adjusted super-Gaussian beam then enters a spatial light modulator, where it
is modulated to impart the desired phase for generating a required vortex
beam. This vortex beam is vertically incident on the input surface of a
nonlinear sample. The nonlinear sample is placed in a high-transmittance glass
tube with a length of $L=13$ cm, filled with a graphene/methanol solution as
the nonlinear medium. The vortex beam output from the nonlinear sample is
transmitted through a $4f$ imaging system ($f=75$ mm, the $4f$ system in our
experiment is not perfect \cite{n4f} and see the appendix for the detail) to a
pinhole, where it is tracked and scanned. The pinhole has a diameter of $200$
$\mu$m and is mounted on a motorized translation stage that allows for
high-precision, three-dimensional directional control. The three-dimensional
high-precision motorized translation stage can accurately control the radial
scanning of the pinhole along the vortex beam, with a unidirectional movement
precision of up to $3$ $\mu$m. Subsequently, a lens with a focal length of
$50.8$ mm is used to perform a Fourier transform on the spot selected by the
pinhole, converting the frequency distribution of the spot into the spatial
distribution. Therefore, the spatial frequency components of the spot can be
observed at the focal plane of the lens. After the Fourier transform by the
lens, a charge-coupled device (CCD) camera located at the focal plane of the
lens records the far-field intensity distribution at the output surface of the
nonlinear sample.

In particular, we use an exposure time of 200 $%
\mu
$s to acquire the near-field intensity distribution and to capture the
far-field point image by focusing the light onto the Fourier plane using a
lens, where the far-field spot is selected through a pinhole. This exposure
time is much shorter than the relaxation time from non-equilibrium to
equilibrium state in the thermo-optic nonlinear process, and our photos are
taken after the system reached a stationary state. Furthermore, because the
spatial extent (less than $166$ $\mu$m) of the nonlocal response function is
much less than the sound length ($\sim1$ mm) in our experiment, the nonlocal
response function can be approximated as local, which supports the
interpretation below Eq. (\ref{index}).

\begin{figure}[ptb]
\centering
\includegraphics[width=0.9\columnwidth]{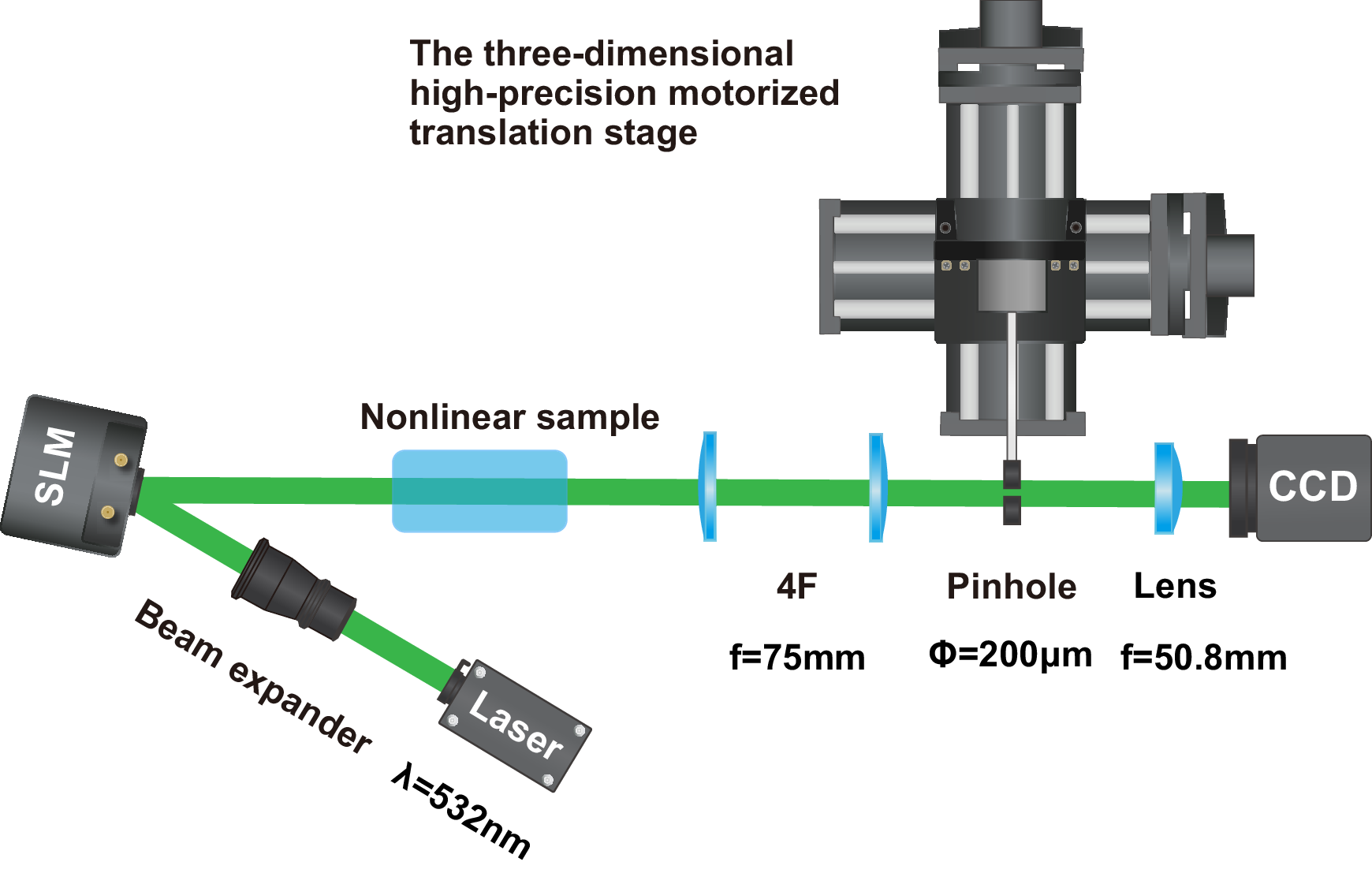} \caption{(Color online)
Experimental setup: a continuous wave $532$ nm laser beam is emitted into a
beam expander for expansion. The expanded beam then enters the SLM (Spatial
Light Modulator) to modulate the vortex beam phase as required. The generated
vortex beam passes through a $13$ cm long glass tube filled with a
graphene/methanol solution and is further expanded using a $4f$ system. The
expanded vortex beam then passes through a pinhole (which selects a radial
region of the vortex beam and focuses it on the focal plane of the lens), and
the spatially resolved far-field is recorded by a CCD camera.}%
\label{Fig2}%
\end{figure}

The photon fluid flow can be measured by recording the spatially resolved
components $K_{x,\text{ }y}(r)$ in the far field, that is $(v_{x},v_{y}%
)=\frac{c}{n_{0}k_{0}}\left(  K_{x},K_{y}\right)  $. In the experiment, the
pinhole's position in the vertical direction is fixed at $y=-0.156$ mm. By
changing its position in the horizontal direction ($x$) and using the pinhole
to track and scan different radial positions of the vortex beam, the location
information of the spots captured by the CCD can be used to calculate the flow
velocity. The necessary measurement for the phonon velocity is the intensity
distribution of the vortex beam after the interaction by taking a photo for
the vortex beam in the near field. In particular, it is stressed that the
measurement of the light intensity is taken in regions very close to the core
and the corresponding results in Fig. 3 are the average values across $x$ axis
at the same $y$ position excluding the dark vortex core. The detailed
measurement method for the velocities are given in the appendix.

\begin{figure}[ptb]
\centering
\includegraphics[width=0.9\columnwidth]{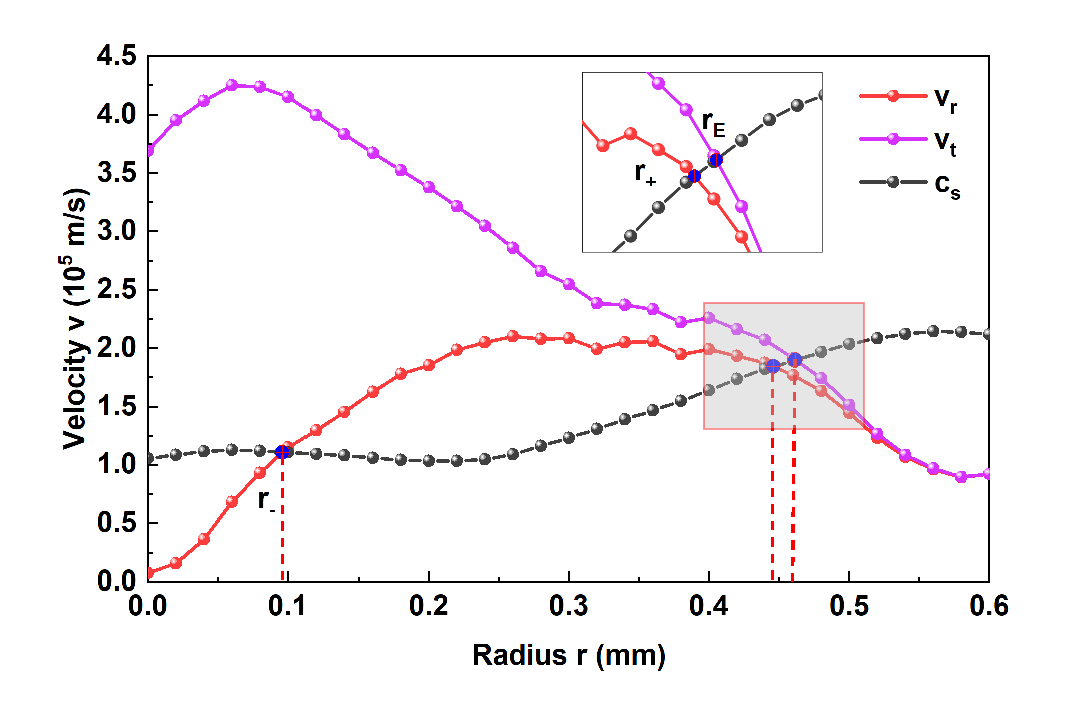} \caption{(Color online)
Experimental data for the velocities of the fluid and the phonon. The red line
represents the radial velocity which leads to the horizons at the
intersections with the black line of the phonon speed. The purple line
represents the total velocity which gives the position for the ergosurface
when it matched with the photon speed. The parameters take the same values as
in Fig. 1, but the power of the laser reduces to $70$ mW due to the nonlinear
interaction. The measurement error is $\pm0.003$ unit for radius in the
abscissa axis and $\pm0.02$ unit for velocities in the vertical axis.}%
\label{Fig3}%
\end{figure}

Figure 3 shows the velocity curves for the photon fluids and the excited
phonons. Acoustic horizons can be identified at the intersections of the fluid
velocity (red line) and the phonon velocity (black line) at $r=0.0962$ mm and
$r=0.4459$ mm, corresponding to the inner horizon (labeled as $r_{-}$) and the
outer horizon (labeled as $r_{+}$), respectively. The ergosurface is larger
than the horizons and is located at $r=0.4616$ mm where the total flow
velocity (purple line) intersects the phonon velocity (black line), labeled as
$r_{E}$. These intersections present evidently the analogous spacetime
structure given by the rotating photon fluid after interaction with the
nonlinear medium. It is noted that Fig. 3 shows good agreement with the
simulation results in Fig. 1, with the specific locations of the three
intersections also satisfying the condition ($r_{-}<r_{+}<r_{E}$).

It is able to estimate by the experimental parameters that the analogous
spacetime is well within the sonic regime by calculating the angular velocity
of the horizon (or the absorbing boundary in the Zel'dovich effect),
$\omega_{Z}\simeq\frac{v_{\theta}(r_{+})}{r_{+}}\simeq1.97\times10^{8}$
s$^{-1}$ which corresponds to phonons with the wavelength $\sim2\pi
c_{s}/\omega_{Z}\simeq4.15$ mm, or the analog surface gravity $\kappa=\frac
{1}{2}\partial_{r}\left(  c_{s}^{2}-v_{r}^{2}\right)  |_{horizon}%
\simeq1.88\times10^{13}$ ms$^{-2}$ which corresponds to phonons with the
wavelength $\sim c_{s}^{2}/\kappa\simeq1.82$ mm. Therefore, we have
successfully simulated the acoustic black hole with its structure analogue to
a rotating BTZ black hole in (2+1) dimensional spacetime using the vortex beam
in the laboratory.

Finally, we have to check whether the analogue is made in the effective regime
with the boundary determined by $R_{\pm}$ in the Eq. (\ref{fbhb}). At first,
we express the mass and angular momentum of the analogous black holes as
\begin{equation}
M=\frac{v_{t}^{2}/c_{s}^{2}-1}{1-r_{+}^{2}/r_{E}^{2}},\text{ }J=\frac
{2r_{+}v_{\theta}}{c_{s}},
\end{equation}
where the radius of anti-de Sitter space given by $r_{E}=\sqrt{M}l$ is used
for the expression of the mass. Thus, all the quantities for the black hole
are determined by the experimental parameters, which leads to the analogue
mass $M\simeq3.07$, $J\simeq4.05\times10^{-4}$ and $l\simeq2.63\times10^{-4}$
(up to $c^{2}/G\sim10^{27}$ kg which is ignored in our discussion). By these,
we calculate $R_{+}\simeq0.52$ mm and $R_{-}\simeq0.10$ mm, which shows the
inner and outer horizons and the ergosurface are well simulated experimentally
in the effective regime.

\section{Conclusion}

In this paper, we experimentally investigate how to use the photon fluid of a
vortex beam to simulate a BTZ black hole, presenting an analogy to a rotating
black hole with a more complete structure. In the experiment, we recorded
videos of the light spot movement through a pinhole during scanning and
extracted key frames to capture the changes in the light spot's position. This
allowed us to analyze the radial flow velocity of the vortex beam and the
movement speed of phonons more efficiently and in greater detail. By plotting
the velocity curves, we clearly identified the positions of the BTZ black
hole's inner horizon, outer horizon, and static boundary, simulating a more
complete spacetime structure of the BTZ black hole. This opens a more
extensive application of the optical vortex and provides a possibility for
further experimental research into the physical properties related to rotating
black holes, such as Hawking radiation, Penrose radiation, and even the
AdS/CFT correspondence.

\section{Acknowledgments}

This work is supported by National Natural Science Foundation of China (NSFC)
with Grant No. 12375057 and the Fundamental Research Funds for the Central
Universities, China University of Geosciences (Wuhan).

\section{Appendix: Experimental method for measurement of velocities}

Here we introduce the method about how to measure the velocities
experimentally. As illustrated in the Fig. 2 of the main text. The vortex beam
output from the nonlinear sample is modulated in its intensity distribution by
the $4f$ system, resulting in the vortex beam shown in the Fig. 4(a), which is
the near-field image and is taken when the pinhole and the lens are removed
and can be used to measure the light intensity distribution. The interaction
between the vortex beam and the nonlinear medium causes the beam's center to
contract. The modulation of the vortex beam by the $4f$ system allows for
easier extraction and identification of more detailed spot information from
its ring structure. In particular, the $4f$ system is not perfect in our
experiment since the first lens is placed at the distance $f$ (=$75$ mm) from
the focus but the second lens is placed at the distance a little longer ($80$
mm) from the focus. This amplifies the image of the vortex to observe and
measure easier, but this does not alter the crucial structure of the vortex
after the nonlinear interaction with the medium. The Fig. 4(b) shows the spot
images captured by the CCD after the pinhole selection and focusing onto the
lens's focal plane, which is the far-field image used for the measurement of
the fluid velocities.

\begin{figure}[ptbh]
\centering
\includegraphics[width=1\columnwidth]{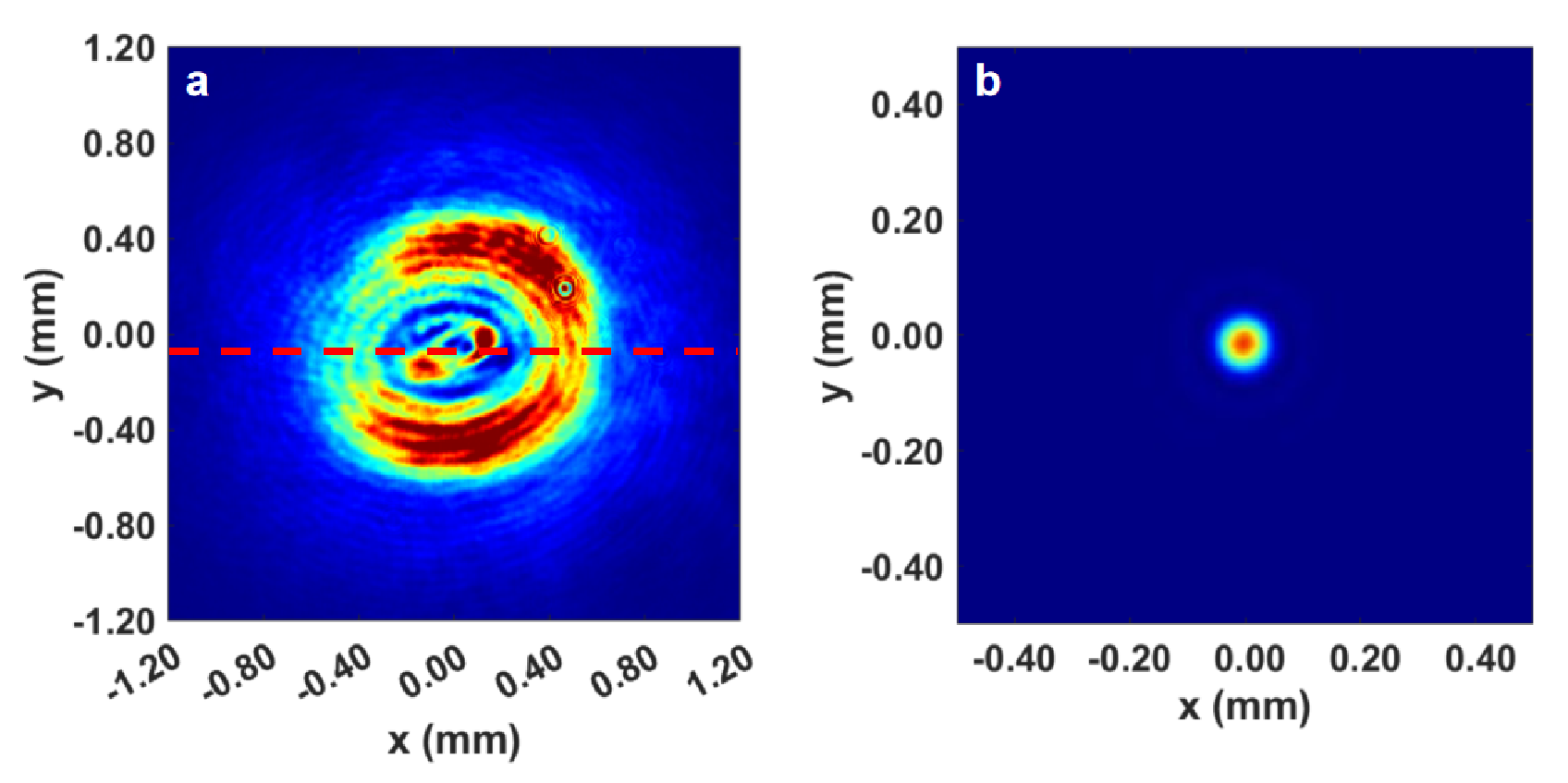}\caption{(Color online) The figure
(a) presents the vortex beam after expansion by the 4f system. The figure (b)
presents the far-field vortex beam at the positions ($K_{x}(x,y=0),K_{y}%
(x,y=0)$). }%
\label{FigS1}%
\end{figure}

Since the velocity is closely related to the nonlinear interaction process, we
carefully control the waiting time and exposure time during experiments to
ensure accurate measurements. For thermal optical nonlinearity induced by
continuous light, the response time depends on the beam intensity (power and
focusing radius). For example, a pump beam with a power of $70$ mw, the
response time can exceed $1$ second. In our setup, we use a continuous laser
source with power stability of more than $99\%$ and allow the vortex beam to
interact with the nonlinear medium continuously without a shutter until a
stable thermal nonlinear state is achieved. Once stable, we record phase and
intensity distributions using a total acquisition time of $5$ seconds, with
each frame exposure time controlled to within $200$ $\mu$s to ensure stable
and precise measurements.

For the measurement of the sound speed, it is crucial for the methanol to
exhibit thermo-optic nonlinearity induced by laser power, characterized by its
nonlinear refractive index, with its value as $\gamma=-4.4\times10^{-11}$
m$^{\text{2}}$/W. During the preparation of the solution, adding an
appropriate amount of nanoscale graphene sheets to methanol can significantly
enhance the absorption efficiency of the sample and provide sufficient
nonlinearity required for experimental research. The key measurement is the
light intensity distribution $I$. In our experiment, multiple regions equal to
the pinhole aperture area along the pinhole's movement path (as indicated by
the dashed lines in the Fig. 4(a)) were selected for grayscale integration
calculations. Subsequently, we compared the grayscale integration value of
each selected region with the total grayscale integration on the CCD detection
plane, thereby quantitatively obtaining the power distribution of the selected
spots at different positions during the pinhole's movement. To analyze the
radial variation characteristics of the light intensity distribution in
detail, the total grayscale integration values of the vortex beam images
captured on the CCD plane were calculated. Along the radial direction
($x$-axis), we calculated the grayscale integration value within the pinhole
aperture every $0.2$ mm. In our experiment, the camera's pixel size is $3.45$
micrometers, which means that for the captured spot images, the grayscale
integration within an area equal to the pinhole aperture was calculated
approximately every $6$ pixels to refine the spatial distribution of light
intensity. This series of detailed calculations not only improved the accuracy
of the data but also allowed us to closely track the spatial variations in
light intensity distribution. Based on these calculations, we successfully
plotted the curve of phonon velocity changes.

It is worth noting that when processing light spot images, the grayscale
values of the images must be taken into account. Therefore, it is essential to
carefully control the camera's exposure time during image capture. Excessively
long exposure times may lead to large areas of overexposure in the vortex beam
image, causing the grayscale values in these areas to exceed the maximum
threshold of $255$, which can adversely affect the accuracy of the
experimental data. Conversely, if the exposure time is too short, the light
spot image may not be adequately exposed, resulting in lower grayscale values
that fail to accurately represent the true brightness of the light spot. In
this case, the captured image will have fewer details, reducing the accuracy
of the analysis and compromising the reliability of the experimental results.
By properly controlling the exposure time, it is possible to ensure that the
light spot images maintain rich detail while accurately reflecting changes in
the light spot's brightness.

For the measurement of the flow speed, we first used a CCD camera to capture
the imaging of the vortex beam focused on the focal plane by the lens before
measurements, without any pinhole filtering. The peak position at the center
of the spot was used as a reference for comparative analysis. Then, using a
three-dimensional high-precision motorized translation stage, we performed
radial scanning at a speed of $0.2$ mm/s, starting from the edge of the vortex
beam. The entire process was video recorded by the CCD camera. We extracted
each frame of the spot images from the video, converted the images to
grayscale after reading the files, and simplified the analysis process. For
the converted grayscale images, we applied image morphology operations to
determine the center and area of each region, selecting the region with the
largest area as the location of the brightest spot. This allowed us to obtain
the position information of the brightness peak at the center of the spot at
different radial coordinates of the vortex beam (after selection by the
pinhole and focusing by the lens). Based on this position information, the
radial flow speed and total flow speed can be calculated.

\begin{figure}[ptbh]
\centering
\includegraphics[width=1\columnwidth]{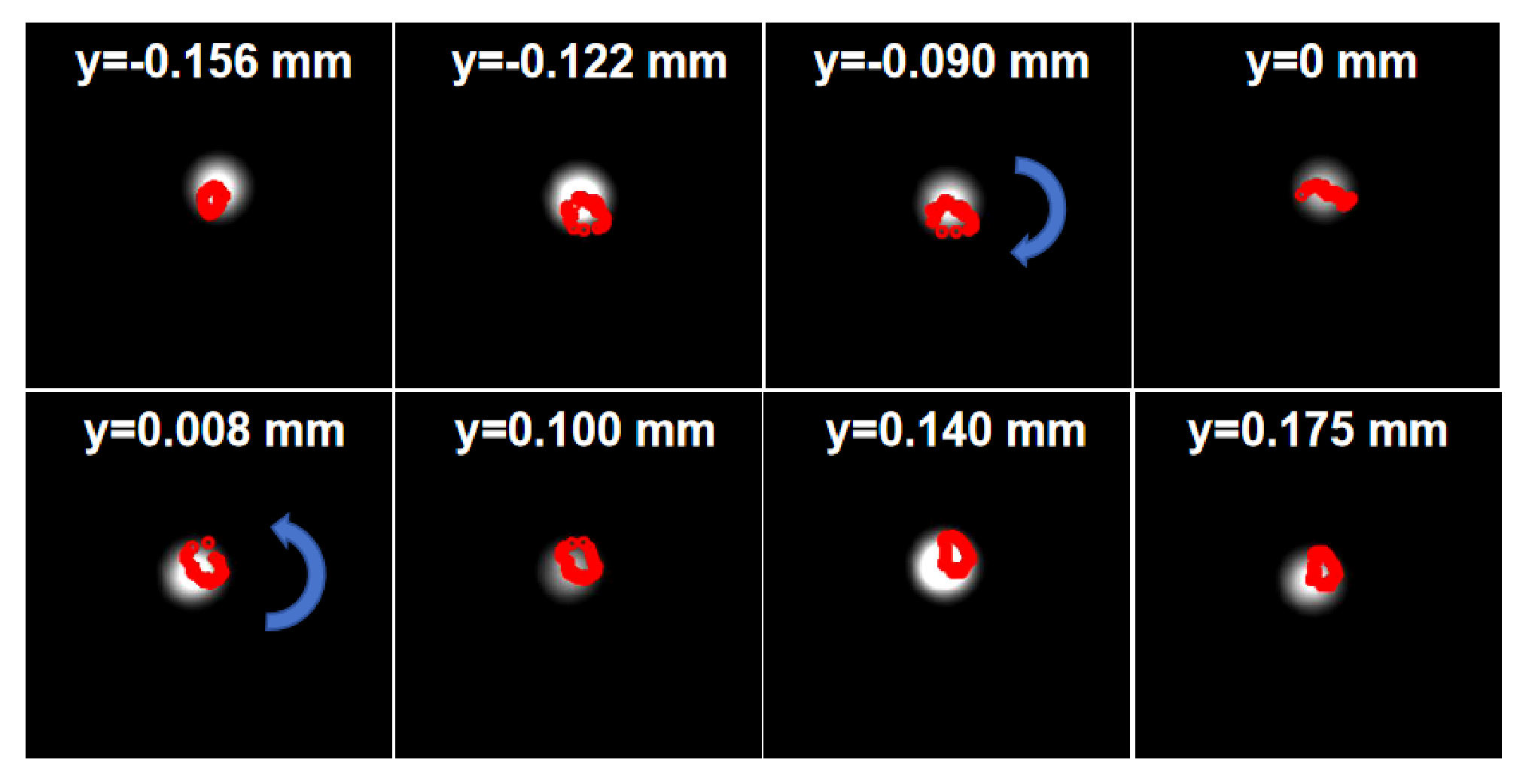}\caption{(Color online) The
trajectories of the light spots obtained by scanning at different longitudinal
positions near the vortex beam center (y=0 mm). Scanning above the vortex beam
center is defined as the positive direction, while scanning below the vortex
beam center is defined as the negative direction. }%
\label{FigS2}%
\end{figure}

It is important to note that in actual optical measurements, due to the zero
light intensity at the center of the vortex beam, if the pinhole path
coincides with the dark core center of the vortex beam during a surface scan,
the light spot trajectory may lose some information as the light intensity
inside the pinhole is lower when passing through the dark core. To address
this issue, we perform scans in areas very close to the vortex beam's center.
We chose to scan above and below the center of the vortex beam repeatedly.
This operation approximates the scenario where the light spot passes through
the vortex beam center, allowing for a more accurate acquisition of the
required optical information, and can be considered as if the measurement
process passes through the optical center. By repeatedly scanning near the
vortex beam center, we can minimize the impact of partial loss of the light
spot trajectory, thereby improving the accuracy and reliability of the measurements.

As shown in Fig. 5, this is a schematic diagram of the light spot trajectory
obtained by scanning at different longitudinal positions near the optical
center ($y=0$ mm). Scanning above the vortex beam center is defined as the
positive direction, while scanning below the vortex beam center is defined as
the negative direction. The direction of movement for the trajectory above the
$y=0$ mm position (position greater than $0$ mm) is opposite to that of the
trajectory below the $y=0$ mm position (position less than $0$ mm). Complete
light spot trajectories can be formed by scanning at certain distances above
and below the optical center. However, after the pinhole passes through the
dark core region at the center of the vortex beam, part of the light spot
trajectory is missing, making it impossible to form a complete closed
trajectory. So, selecting a closed trajectory is important for measuring the
velocities successfully. Below, we will conduct a detailed analysis of the
closed trajectory at the position $y=-0.156$ mm.

\begin{figure}[ptbh]
\centering
\includegraphics[width=1\columnwidth]{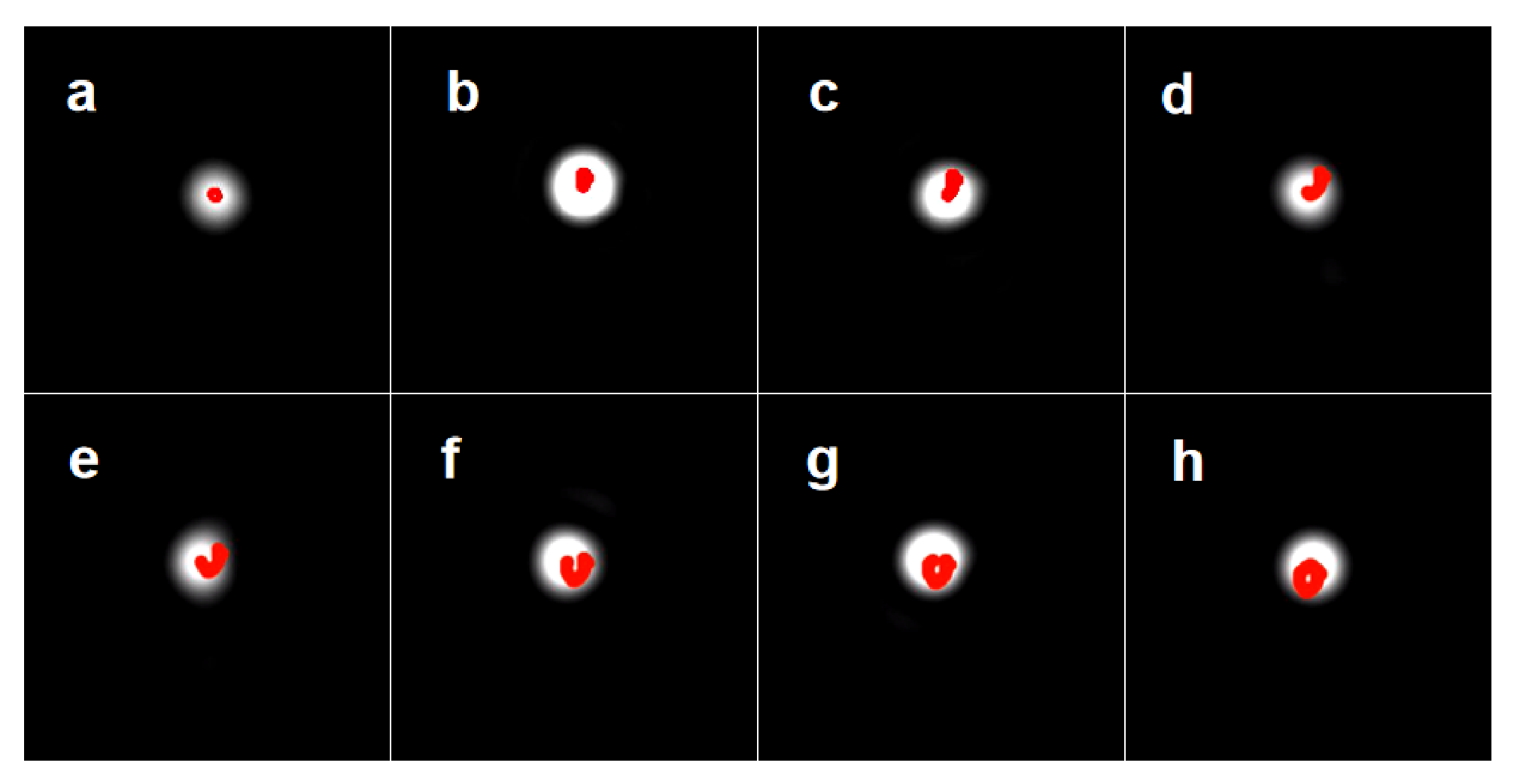}\caption{(Color online) The
trajectory for the light spot at a distance of y = -0.156 mm from the vortex
beam center. }%
\label{FigS3}%
\end{figure}

Figure 6 shows the partial light spot trajectory at the position $y=$ $-0.156$
mm. Figs. 6(a) to 6(h) illustrate the movement of the light spot center as the
pinhole moves radially along the vortex beam. Before the scanning begins, the
center of the light spot selected by the pinhole is located at the position
shown in Fig. 6. As the pinhole starts moving radially along the vortex beam,
the position of the light spot also begins to change, with the optical center
gradually moving in a clockwise direction, eventually forming a closed
trajectory. It is important to note that by the end of the scan, the light
spot returns to its initial position, coinciding with the light spot center
before the scan began, as shown in Fig. 3h.


\begin{thebibliography}{99}
\bibitem {n4f}The non-perfect $4f$ system causes only a minor modification for
the measurement results, which even cannot be presented clearly in the figure
of the experimental results. As well-known, when the light goes through the
lens, the phase is modulated with a factor, $\varphi_{m}=\frac{k}{2f}\left(
x_{f}^{2}+y_{f}^{2}\right)  $, which leads to a modification for the phase
with $\Delta\varphi=\frac{k}{2f}\left(  1-\frac{d}{f}\right)  \left(
x_{f}^{2}+y_{f}^{2}\right)  =\left(  1-\frac{d}{f}\right)  \varphi^{\prime
}=\frac{1}{15}\varphi^{\prime}$ where $d$ is the distance between the focus of
the first lens and the second lens, and $\varphi^{\prime}$ is the phase that
the light carries after going through the second lense which is placed
according to the perfect $4f$ system. When taking the derivative of the phase
to obtain the velocity, the modification contributes too small to present in
the figure clearly.
\end{thebibliography}
\end{document}